\def \PR {{ Phys. Rev.} }

\def \NP {{ Nucl. Phys.} }
\def \PRL {{ Phys. Rev. Lett. }}
\def \bc {\begin{center}}
\def \ec {\end{center}}

\def \bfr {\begin{flushright}}
\def \efr {\end{flushright}}

\def \ba {\begin{array}}
\def \ea {\end{array}}

\def \bea {\begin{eqnarray}}
\def \eea {\end{eqnarray}}

\def \be {\begin{equation}}
\def \ee {\end{equation}}

\def\tr {\,{\rm tr}}
 
\def\overpmup {\overleftarrow{\partial}\,\,\!\!\!\!\!\!\!
\overrightarrow{\hbox{\ \ }^\mu}}

\def\ni{\noindent}
\def\nn{\nonumber}
\def\p{\partial}
\def\f{\frac}
\def\l[{\left[}
\def\r]{\right]}
\def\TG{\tilde{G}}
\def\TT{\tilde{T}}

\def\tg{\tilde{g}}
\def\tk{\tilde{g_t}}
\def\tp{\tilde{g}_p}

\def\tit{\tilde{g}_t}
\def\um{\frac{1}{2}}
\def\c{\varphi_{{}_+}}
\def\cc{\varphi_{{}_-}}
\def\bos{Q}
\def\bosc{\bar{Q}}
\def\dos{D}
\def\dosc{\bar{D}}
\def\ch{\chi_{{}_+}}
\def\chc{\chi_{{}_-}}
\newcommand{\deriv}[1]{ \frac{\delta}{\delta #1} }
\newcommand{\coci}[1]{ \frac{i}{r^2}\int{{d}^3x\,\tr\left[ #1 \right]}}
\newcommand{\xl}[1]{ {\tilde{X}}^{L}_{#1} }

\newcommand{\xr}[1]{ {\tilde{X}}^{R}_{#1} }

\documentstyle[12pt,a4]{article}

\begin{document}

\begin{flushright}
{\large SWAT-99/224}
\end{flushright}

\begin{center} 
{\Large {\bf Group Approach to Quantization of Yang-Mills Theories: A 
Cohomological Origin of Mass}}
\footnote{Work partially
supported by the DGICYT.}
\end{center}
\bigskip
\bigskip
\centerline{ {\it Manuel Calixto$^{1,3}$\footnote{E-mail: pymc@swansea.ac.uk
/  calixto@ugr.es}  
and   V\'\i ctor Aldaya$^{2,3}$\footnote{E-mail: valdaya@iaa.es}} }
\bigskip

\begin{enumerate}
\item {Department of Physics, University of Wales Swansea, Singleton Park, 
Swansea, SA2 8PP, U.K.}
\item {Instituto de Astrof\'{\i}sica de Andaluc\'{\i}a, Apartado Postal 3004,
18080 Granada, Spain.}
\item  {Instituto Carlos I de F\'\i sica Te\'orica y Computacional, Facultad
de Ciencias, Universidad de Granada, Campus de Fuentenueva, 
18002 Granada, Spain.} 
\end{enumerate}

\bigskip
\begin{center}
{\bf Abstract}
\end{center}
\small

\begin{list}{}{\setlength{\leftmargin}{3pc}\setlength{\rightmargin}{3pc}}
\item 
New clues for the best understanding of the nature of the symmetry-breaking 
mechanism are revealed in this paper. A revision of 
the standard gauge transformation properties of Yang-Mills 
fields, according to a group approach to quantization scheme, enables the 
gauge group coordinates to acquire dynamical content outside 
the null mass shell. The corresponding extra (internal) 
field degrees of freedom are transferred to the vector potentials to  
conform massive vector bosons. 
\end{list}

\normalsize
\noindent PACS: 11.15.-q, 03.65.Fd, 02.20.Tw, 11.15.Ex\\
KEYWORDS: gauge theories, symmetry breaking, groups, cohomology, 
algebraic quantization. 
\vskip 1cm
\section{Introduction}

Despite the undoubted success of the Standard Model in describing strong 
and electro-weak interactions, a {\it real} (versus artificial) 
mechanism of mass-generation is still lacking. Needless to say that the 
discovery of a Higgs boson ---a quantum vibration of an {\it abnormal} (Higgs) 
vacuum--- would be of enormous importance; 
nevertheless, at present, no dynamical 
basis for the Higgs mechanism exists, and it is purely phenomenological. 
It is true that there is actually nothing inherently 
unreasonable in the idea that the 
state of minimum energy $|\tilde{0}\rangle$ (the vacuum) 
may be one in which some field quantity 
$\hat{\varphi}(x)$ has a non-zero expectation value $\langle \tilde{0}|
\hat{\varphi}(x)|\tilde{0}\rangle=\varphi_0$; in fact, 
many examples in condensed-matter physics display this feature. Nevertheless, 
it remains conjectural whether something similar actually happens in the weak 
interaction case.

Also, the {\it ad hoc} introduction of extra (Higgs) scalar fields 
in the theory to provide mass to the 
vector bosons could be seen as our modern equivalent of those earlier 
mechanical contrivances populating the {\it plenum} (the ether), 
albeit very subtly. As in those days, new perspectives are necessary 
to explain why it is really not indispensable to look at things 
this way at all. 

This paper provides a new approach to quantum Yang-Mills  theories, from 
a group-theoretic perspective, in which mass enters the theory 
in a {\it natural} way; more precisely, the presence of mass will manifest 
through non-trivial transformations of the phase $\zeta=e^{i\alpha}$ 
of the wave functional $\Psi(A)$ under the action of gauge transformations. 
This non-trivial response of the phase under gauge transformations 
causes a {\it deformation} of the corresponding 
Lie-algebra commutators and leads to 
the appearance of central terms proportional to  mass parameters and, 
consequently,  to 
a quantum generation of extra (internal) field degrees 
of freedom according to a self-interacting theory of massless 
and massive vector bosons (without Higgs fields). 

This {\it cohomological} mechanism of mass-generation makes perfect sense 
from a Group Approach to Quantization (GAQ \cite{GAQ}) framework, and 
we shall use its concepts and tools to work out the quantization 
of Yang-Mills theories. Given that this is not a common 
approach to quantization, we shall give useful references and 
try to be as self-contained as possible (the reader is advised to have a look 
at the Ref. \cite{config}, which contains a general  
presentation of GAQ for linear fields). Quantizing on a group, however,  
will require the revision of some standard concepts,  
such as {\it gauge transformations}, in order to deal with them properly. 
The meaning of gauge transformations in Quantum Mechanics is not well 
understood at present (see, for example, \cite{Rovelli}); thus, a 
re-examination of it is timely.  

Gauge symmetry is always a guarantee for the renormalizability of a field  
theory. The introduction of mass usually spoils gauge invariance, but 
the Higgs mechanism manages to preserve renormalizability by 
keeping gauge invariance in a {\it hidden} way, and this is the main 
novelty in comparison with other attempts to supply mass. 
However, we must say that the breakdown 
of a gauge symmetry and the appearance of anomalous (unexpected) situations 
are sometimes  subtle questions which generally go with 
the standard approach of {\it quantizing classical systems}. 
From a group-theoretic framework, any consistent 
(non-perturbative) quantization is just a unitary irreducible representation 
of a suitable (Lie, Poisson) algebra. This approach does not assume 
the existence of a previous classical underlying system and 
overcomes some of the standard failures in quantization 
(anomalies) attached to canonical quantization, reinterpreting them as 
normal (even essential) situations.

A unified quantization of massless and massive non-Abelian vector bosons 
will be presented in Sections \ref{yms} and \ref{ymc}, respectively;  
the Abelian case (Electromagnetic and Proca fields) 
is briefly discussed in Sec. \ref{Abelian}. 
The Hilbert space of the theory is related to the carrier space 
of the unitary irreducible representations of an 
infinite-dimensional {\it quantizing group} $\TG$, the mass eventually 
being a parameter characterizing the representation. Section 
\ref{fermionic} is devoted 
to the incorporation of fermionic matter into the theory. Finally, we conclude 
in Section \ref{conclu} and incorporate an Appendix with a simple, 
but illustrative, 
finite-dimensional analogy.

\section{The Abelian case \label{Abelian}}

In a previous article \cite{empro},  a revision of the traditional concept 
of gauge transformation for the electromagnetic vector potential, 
\be 
\varphi(x)\rightarrow \varphi'(x)+\varphi(x),\;\;\;
A_{\mu}(x)\rightarrow A_{\mu}(x)-\partial_{\mu} \varphi'(x)\,,
\label{connection} 
\ee
\ni was necessary to arrange this transformation inside a group law that is, 
to adapt this operation to an action of a group on itself: 
the group law of the (infinite-dimensional) 
{\it electromagnetic quantizing group} $\TG$. The proposed Lie group $\TG$
had a principal bundle structure $\TG\rightarrow\TG/\TT$ and was
parameterized, roughly speaking, by the coordinates $A_{\mu}(\vec{x},t)$ of
the Abelian subgroup $G_A$ of Lie algebra valued vector potentials, 
the coordinates $v=(y_\mu,\Lambda_{\mu\nu})$ (space-time translations 
and Lorentz transformations) of the Poincar\'e group $P$ and the coordinates 
$\varphi(\vec{x},t)$ of the local group
$T\equiv U(1)(\vec{x},t)$, which took part of the structure group 
$\TT\sim T\times U(1)$ and 
generalized the standard $U(1)$-phase 
invariance, $\Psi\sim e^{i\alpha}\Psi$, in Quantum Mechanics. 
In this way, the extra
$\TT$-equivariance conditions on wave functions [complex valued functions 
$\Psi(\tg)$ on $\TG$], i.e. 
$\Psi(\tg_t*\tg)\sim\Psi(\tg),\,\,\tg_t\in  \TT$, 
provided the traditional constraints of the theory.

The abovementioned revision 
was motivated by the fact that the transformation (\ref{connection}) 
is not compatible with a group law.  Indeed, the general property
$g*e=e*g=g$ for a composition law $g''=g'*g$ of a group $G$ ($e$ denotes
the identity element), precludes the existence of linear terms, 
in the group law $g''^j=g''^j(g'^k,g^l)$ of a 
given  parameter $g^j$ of $G$, 
other than $g'^j$ and $g^j$; that is, near the identity 
we have $g''^j=g'^j+g^j+O(2)$. Therefore, the group law for 
the field parameter $A_{\mu}$ cannot have linear terms in $\varphi$. 
The natural way to address this situation is to 
leave the vector potential unchanged, and change 
the phase $\zeta=e^{i\alpha}$ 
of the quantum-mechanical wave functional $\Psi(A)$ as follows:
\bea
\varphi(x)\rightarrow \varphi(x)+\varphi'(x),\;\;\;
A_\mu(x)\rightarrow A_\mu(x),\nn \\
\zeta\rightarrow \zeta 
\exp\left\{-\frac{i}{2c\hbar^2}\int_\Sigma d\sigma_\mu(x) 
\eta^{\rho\sigma}\partial_\rho \varphi'(x)\overpmup 
A_\sigma(x)\right\}\, ,\label{gaugenew} 
\eea
\ni where  $\eta^{\rho\sigma}$ denotes the Minkowski metric, 
$\Sigma$ denotes a spatial hypersurface and $\hbar$ is the Planck constant,
which is required to kill the dimensions of $\partial_\rho\varphi'\overpmup 
A^\rho\equiv\partial_\rho\varphi'
\partial_\mu  A^\rho- A^\rho
\partial_\mu\partial_\rho \varphi'$ and gives a 
{\it quantum} character to the transformation (\ref{gaugenew}) versus 
the {\it classical} character of (\ref{connection}) [hereafter, we shall use 
natural unities $\hbar=1=c$]. The piece  $\partial_\rho\varphi'\overpmup 
A^\rho$  takes part of a {\it symplectic} current 
\be
J^\mu(g'|g)(x)\equiv\um\eta^{\rho\sigma}[(vA')_\rho(x)-
\partial_\rho(v\varphi')(x)]\overpmup [A_\sigma(x)-
\partial_\sigma\varphi(x)]\,, 
\ee
[we are denoting $g\equiv(A,\varphi,v)$ and $(vA')_\rho(x)\equiv 
\f{\p v^\alpha(x)}{\p x^\rho}A'_\alpha(v(x))$, $(v\varphi')(x)\equiv 
\varphi'(v(x))$,  with 
$v^{\alpha}(x)=\Lambda^{\alpha}_\beta x^\beta +y^\alpha$ the general 
action of the restricted Poincar\'e group $P$ on Minkowski space-time] 
which is conserved,
$\partial_\mu J^\mu=0$,  if  $A_\nu$ and $\varphi$ satisfy the field
equations $(\partial_\mu\partial^\mu + m^2) A_\nu=0$ and
$(\partial_\mu\partial^\mu + m^2) \varphi=0$ ($m$ is a parameter with mass
dimension), so that the integral in (\ref{gaugenew}) does not depend on the
chosen space-like hypersurface $\Sigma$. The integral $\xi(g'|g)\equiv
\int_\Sigma d\sigma_\mu(x)J^\mu(g'|g)(x)$  is a two-cocycle $\xi: G\times
G\rightarrow \Re$ [$G$ denotes the semi-direct product 
$(G_A\times T)\times_v P$],  which fulfis the well-known properties:
\be\ba{cc}
\xi(g'|g)+\xi(g'*g|g'')=\xi(g'|g*g'')+\xi(g|g'')\,,\;\;\;
\forall g,g',g''\in G\,,\\
\xi(g|e)=0=\xi(e|g)\,,\;\;\;\forall g\in G\,, \ea
\ee
\ni and is the basic ingredient to construct the 
centrally extended group law $\tg''=\tg'*\tg$, more explicitly
\be 
\tg''\equiv (g'';\zeta'')=(g'*g;\zeta'\zeta e^{i\xi(g'|g)})\,,
\;\;\; g,g',g''\in G;\,\,\zeta, \zeta',\zeta''\in U(1)\,,\label{leyext}
\ee
of the  electromagnetic quantizing group $\TG$ (see below and Ref.
\cite{empro} for more details). 

It bears mentioning that the required revision of the concepts 
of gauge transformations and constraint 
conditions to construct the quantizing group $\TG$ has led, as a byproduct, 
to a unified quantization of both the electromagnetic and Proca fields 
\cite{empro}, within the same general scheme of 
quantization based on a group (GAQ) \cite{GAQ}. The different structure 
of the central extension (\ref{leyext}) for the massive case, with 
regard the massless case, manifests itself through a {\it true} (non-trivial) 
central-extension $\TT$ of the constraint subgroup $T$ by $U(1)$ 
given by the  peace 
\be
\xi_m(g'|g)=\um\int_\Sigma{ d\sigma_\mu \eta^{\rho\sigma}
\partial_\rho(v\varphi') \overpmup \partial_\sigma\varphi}=
\frac{m^2}{2}\int_\Sigma{ d\sigma_\mu (v\varphi')\overpmup\varphi}\,,
\label{coci3}
\ee
of the cocycle $\xi(g'|g)\equiv
\int_\Sigma d\sigma_\mu J^\mu(g'|g)$. The piece $\xi_m$, which 
is one ($\xi_3$) of the three typical and distinguishable pieces 
($\xi_j,\, j=1,2,3$) in which $\xi$ splits up (see \cite{empro,gtp} and 
bellow), gives dynamics to the local group $T$ (creates new couples of 
conjugated variables), and makes the constraints 
of second-class nature. This results in an increased  number of field degrees 
of freedom with regard the massless case, leading to 
a Proca quantum field (see \cite{empro} for more details). 

Furthermore, 
the standard (classical) transformation  (\ref{connection}) is regained 
as the trajectories associated with  generalized equations 
of motion generated by vector fields with null Noether invariants 
({\it gauge subalgebra}, see Refs. \cite{empro,config} and \cite{gtp} for 
a formal exposition, including tensor fields). 

A unified scheme of quantization for non-Abelian massless and 
massive vector bosons is also possible in this scheme and 
suitable as an alternative to the standard 
Spontaneous Symmetry Breaking mechanism, which is intended to supply mass
while preserving renormalizability. 
However, for this case, the situation seems to be a 
bit more subtle and far richer.

\section{Group-quantization of Yang-Mills Fields\label{yms}}

According to the headings in the foregoing section, our main purpose
now is to offer a reasonable attempt to find a  
(non-perturbative) unified quantization of non-Abelian massless and massive 
vector bosons without Higgs fields. As in the Abelian case, the key to 
achieve this goal 
consists in a revision of the traditional concept of gauge transformation 
for vector potentials,
\be
U(x)\rightarrow U'(x)U(x)\,,\;\;\;\;A_\nu(x)\rightarrow 
U'(x)A_\nu(x)U'(x)^{-1}+U'(x)\partial_\nu U'(x)^{-1},\label{transf1}
\ee
in order to make it compatible with a group law: the group law of the 
(infinite-dimensional) {\it Yang-Mills quantizing group} $\TG$, which 
will be the primary object to define the quantum theory. This group has  
a fibre-bundle structure $\TG\rightarrow\TG/\TT$ and is  
parametrized, roughly speaking, by the coordinates 
$A^\mu(x)= r_a^b A^\mu_b(x) T^a$ of an Abelian subgroup $G_A$ of 
Lie algebra valued vector potentials [$r_a^b$ is a 
coupling-constant matrix and $T^a$ are the Lie-algebra generators of 
the rigid subgroup ${\bf T}$, of a gauge group $T$, satisfying the 
commutation relations $[T^a,T^b]=C^{ab}_c T^c$ and defining the structure 
constants $C^{ab}_c$] and the coordinates $U(x)=e^{\varphi_a(x)T^a}\equiv
e^{\varphi(x)}$ of the {\it local} group $T$, which takes part of the 
structure subgroup $\TT\sim T\times U(1)$ and 
generalizes the standard $U(1)$-phase invariance 
$\Psi\sim e^{i\alpha}\Psi$ in Quantum Mechanics as a particular case of  
{\it $\TT$-equivariance} condition \cite{Ramirez}
\bea
\Psi(\tit *\tg)=
D_{\TT}^{(\epsilon)}(\tit)\Psi(\tg)\,,\;\;\forall 
\tit\in \TT,\,\,\forall
\tg\in \TG\,,\label{tequiv}
\eea
on complex wave functionals $\Psi: \TG\rightarrow C$ defined on $\TG$, where 
$D_{\TT}^{(\epsilon)}$ symbolizes a specific 
representation $D$ of $\TT$ with 
$\epsilon$-index (in particular, the $\epsilon=\vartheta$-angle 
\cite{Jackiwtheta} of non-Abelian gauge theories; see below). As already 
commented, 
the $\TT$-equivariance conditions 
(\ref{tequiv}) provide the traditional {\it constraints} of the theory, which 
will be first- or second-class depending on whether the fribration 
of the structure subgroup $\TT\rightarrow \TT/U(1)$ by $U(1)$ is 
trivial or not ($m= 0$ or $m\neq 0$, respectively; see below).

As mentioned above in the Abelian case, 
the transformation (\ref{transf1}) is not 
compatible with a group law. 
The natural way to adapt the operation (\ref{transf1}) to an 
action of a group on itself is to
consider that $A_\nu$  transforms homogeneously under the adjoint action of 
$T$, whereas the non-tensorial part $ U(x)\partial_\nu U'(x)^{-1}$  
modifies the phase $\zeta=e^{i\alpha}$ of the wave functional 
$\Psi(A)$ according to:
\bea
U(x)\rightarrow U'(x)U(x)\,,\;\;\;\;A_\nu(x)\rightarrow 
U'(x)A_\nu(x)U'(x)^{-1}\, ,\nn\\
\zeta\rightarrow \zeta \exp\left\{\frac{i}{r^2}
\int_\Sigma{ d\sigma_\mu(x) 
\tr\left[U'(x)^{-1}\partial_\nu U'(x)\overpmup A^\nu(x)\right]}\right\}\,. 
\label{connection2}
\eea  
We are restricting ourselves, 
for the sake of simplicity, 
to gauge groups $T$ associated with rigid special unitary groups ${\bf T}$ 
for which the structure constants $C^{ab}_c$ are 
totally anti-symmetric, and the anti-hermitian generators $T^a$ can be chosen 
such that the Killing-Cartan metric is just 
${\rm tr}(T^aT^b)=-\um\delta^{ab}$. For simple groups, 
the coupling-constant matrix 
$r_a^b$ reduces to a multiple of the identity $r_a^b=r\delta_a^b$, 
and we have $A^\mu_a=-\f{2}{r}\tr(T^a A^\mu)$. The argument  
of the exponential in (\ref{connection2}) can be considered to be
a piece of a two-cocycle $\xi:G\times G\rightarrow \Re$ 
($G$ is the semi-direct product $G=G_A\times_U T$ of $T$ and the Abelian group 
$G_A$ of Lie-algebra valued potentials) 
constructed through a conserved current,  
$\xi(g'|g)=\int_\Sigma{ d\sigma_\mu(x) 
J^\mu(g'|g)(x)},\,\,\, g',g\in G$, so that it does not depend on the 
chosen spacelike hypersurface $\Sigma$ (see \cite{config,gtp}). 
On this basis, let us 
construct a central extension $\TG$ of $G$ by making use of a two-cocycle
defined on the particular $t={\rm constant}$ 
$\Sigma$-hypersurface. We shall also make 
partial use of the gauge freedom to set the temporal component $A^0=0$, so 
that the electric field is simply $\vec{E}_a=-\partial_0\vec{A}_a$ 
[from now on, 
and for the sake of simplicity,  we shall put any three-vector $\vec{A}$ as 
$A$, and understand $AE=\sum_{j=1}^3A^jE^j$, in the hope that no 
confusion will arise]. In this case, there is still a residual gauge 
invariance $T={\rm Map}(\Re^3,{\bf T})$ (see \cite{Jackiw}). 

The explicit group law 
$\tg''=\tg'*\tg$ [with $\tg=(g;\zeta)=(A,E,U;\zeta)$] for the proposed 
infinite-dimensional {\it Yang-Mills quantizing group} $\TG$ is:

\bea
U''(x)&=&U'(x)U(x)\,,\nn\\
A''(x)&=&A'(x)+U'(x)A(x)U'(x)^{-1}\,,\nn\\
E''(x)&=&E'(x)+U'(x)E(x)U'(x)^{-1}\,,\nn\\
\zeta''&=&\zeta'\zeta\exp\left\{-\frac{i}{r^2}\sum_{j=1}^3
\xi_j(A',E',U'|A,E,U)\right\}\,;
\label{ley}\\
\xi_1(g'|g)&\equiv& \int{{d}^3x\,\tr\left[\,\left(\ba{cc} A' & 
E'\ea\right) S \left(\ba{c} U'AU'^{-1} \\ 
U'EU'^{-1} \ea\right)\right]}\,,\nn\\
\xi_2(g'|g)&\equiv& \int{{d}^3x\,\tr\left[\,\left(\ba{cc} \nabla U'U'^{-1} & 
E'\ea\right) S \left(\ba{c} U'\nabla UU^{-1}U'^{-1} \\ 
U'EU'^{-1} \ea\right)\,\right]}\,,\nn\\ 
\xi_3(g'|g)&\equiv& -2\int{{d}^3x\,\tr\left[ \lambda\left(\log(U'U)
-\log U'-\log U\right)\right]}\,,\nn
\eea
\ni where $S=\left(\ba{cc} 0 & 1 \\ -1 & 0\ea\right)$ is a symplectic 
matrix and $\lambda\equiv \lambda_aT^a$ is a linear function (a matrix) on the
Cartan  subalgebra of the rigid subgroup ${\bf T}$ of $T$.    

We have split up the two-cocycle $\xi$ into three significantly 
distinguishable two-cocycles $\xi_j,\,\,j=1,2,3$ (as in \cite{empro,gtp}) 
for a much better understanding. The first two-cocycle  $\xi_1$ is meant to 
provide {\it dynamics} to the vector potential, so that the couple $(A,E)$ 
corresponds to a canonically-conjugate pair of coordinates. 
The second two-cocycle $\xi_2$, 
the {\it mixed} two-cocycle, provides a non-trivial (non-diagonal) action 
of the structure subgroup $T$ on vector potentials and determines 
the number of degrees of freedom of the constrained theory; it is 
the non-covariant analogue of the argument of 
the exponential in (\ref{connection2}). Both cocycles 
correspond to the analogous ones of the Abelian case. Concerning 
the third one, $\xi_3\equiv\xi_\lambda$, 
its origin and nature departs essentially from the Abelian ``analogue'' 
(\ref{coci3}). Unlike the Abelian case $T=U(1)(x)$, the 
semi-simple character of ${\bf T}$ precludes a {\it true}  
central extension $\TT$ of $T={\rm Map}(\Re^3,{\bf T})$ 
by $U(1)$ (this is not the case in one compact spatial dimension 
$\Re^3\leftrightarrow S^1$, where 
true central extensions are known for Kac-Moody groups). However,
there exists certain coboundaries, called {\it pseudo-cocycles}, which 
define trivial extensions as such, but provide new commutation relations 
in the Lie algebra of $\TG$ and provide a non-trivial piece of the 
{\it connection form} of the theory \cite{GAQ}, 
\be 
\Theta=\left.\f{\p}{\p g^j}\xi(g'|g)\right|_{g'=g^{-1}}dg^j
-i\zeta^{-1}d\zeta\,,\label{thetagen}
\ee
thus altering, in particular, the number of degrees of 
freedom of the theory (see \cite{Marmo} for a relationship between 
{\it pseudo-cohomology} and {\it coadjoint orbits} of semisimple groups). 
This is precisely the case of the third cocycle
(coboundary, indeed), $\xi_3(g'|g)=\eta(g'*g)-\eta(g')-\eta(g)$, 
generated by a function $\eta(g)=-2\int{{d}^3x\,\tr\left[ \lambda\log U
\right]}$  with non-trivial gradient $\left.\delta\eta(g)\right|_{g=e}=
\left.\frac{\delta\eta(g)}{\delta g^j}\right|_{g=e}\delta g^j\not=0$
at the identity $g=e$, which is locally linear in the parameters of the Cartan
subgroup with as many independent coefficients (constants) $\lambda_a$ as 
elements in the Cartan subalgebra, 
i.e. the range of the rigid group ${\bf T}$. 
The introduction of such a pseudo-cocycle is needed to obtain a faithful 
representation of the rigid subgroup ${\bf T}$, according to our general group 
representation approach. Pseudo-cocycles similar to $\xi_3$ do appear in the
representation of Kac-Moody groups and in 
conformally invariant theories in general, 
although the pseudo-cocycle parameters are usually hidden in a redefinition of 
the generators involved in the pseudo-extension (the argument of the 
Lie-algebra generating function). This is the case of the 
Virasoro algebra in String Theory, 
\be
[\hat{L}_n,\hat{L}_m]=(n-m)\hat{L}_{n+m}+
1/12(cn^3-c'n)\delta_{n,-m}\hat{1}\,,\label{viral}
\ee 
where the $\hat{L}_0$ generator is 
redefined so as to produce a non-trivial expectation value in the vacuum,  
$h\equiv (c-c')/24$ \cite{virazorro}.

The cocycle $\xi_3$, however, for $\lambda\not=0$, again determines the 
structure of constraints (first- or second-class) and modifies the dynamical 
content of the vector potential coordinates $A$ by transferring degrees of 
freedom between the $A$ and $\varphi$ coordinates. As in the Abelian case, 
this mechanism conforms massive vector bosons so that $\xi_3$ must be 
considered as a {\it mass cocycle}. In this way, the appearance of mass in 
the theory has a {\it cohomological origin}. Notice that the parameter 
$\lambda$ ($\lambda_a$) bears the dimensions of cubed mass 
(in natural unities) and can well be renamed by $m^3$ ($m_a^3$) . 

To make more explicit the intrinsic 
significance of these three quantities $\xi_j\,,\,\, j=1,2,3$, let us 
calculate the non-trivial Lie-algebra commutators of the right-invariant 
vector fields (that is, the generators of the left-action 
$L_{\tg'}(\tg)=\tg'*\tg$ of $\TG$ on itself) from  the group law (\ref{ley}). 
They are explicitly:
\bea
\l[ \xr{A_a^j(x)}, \xr{E_b^k(y)}\r]&=&-\delta^{ab}\delta_{jk}
\delta(x-y)\Xi\,,\nn\\ 
\l[ \xr{E_a(x)}, \xr{\varphi_b(y)}\r]&=&-C^{ab}_c\delta(x-y)\xr{E_c(x)}+ 
\f{1}{r}\delta^{ab}\nabla_x\delta(x-y)\Xi\,,\label{commutators}\\ 
\l[ \xr{A_a(x)}, \xr{\varphi_b(y)}\r]&=&-C^{ab}_c\delta(x-y)\xr{A_c(x)}\nn\\ 
\l[ \xr{\varphi_a(x)}, \xr{\varphi_b(y)}\r]&=& 
-C^{ab}_c\delta(x-y)\xr{\varphi_c(x)}
-C^{ab}_c\frac{\lambda^c}{r^2}\delta(x-y)\Xi\,,\nn 
\eea 
where we denote by 
$\Xi\equiv i\xl{\zeta}=i\xr{\zeta}$ the central generator, 
in order to distinguish it from 
the rest, in view of its crucial 
role in the quantization procedure; it   
behaves as $i$ times the identity operator, $\Xi\Psi(\tg)=i\Psi(\tg)$, 
when the $U(1)$ part of the 
$\TT$-equivariance conditions (\ref{tequiv}), 
$D_{\TT}^{(\epsilon)}(\zeta)=\zeta$ 
(always faithful, except in the classical limit $U(1)\rightarrow \Re$ 
\cite{GAQ}), is imposed. The commutators (\ref{commutators}) 
agree with those of Ref. \cite{Jackiw} 
when $\lambda_c=0$ and the identification 
$\hat{E}_a\equiv i\xr{A_a}\,,\,\,\hat{A}_a\equiv
i\xr{E_a}\,,\,\,\hat{G}_a\equiv i\xr{\varphi_a}$  is made [note that 
$\xr{A_a}\sim \f{\delta}{\delta A_a}$ and $\xr{E_a}\sim
 \f{\delta}{\delta E_a}$ near the identity element $\tg=e$ of $\TG$, which
motivates this particular identification]. From the last line of 
(\ref{commutators}) we realize that the  pseudo-cocycle $\xi_3$ introduces 
new central terms proportional to the mass parameters $\lambda_c=m^3_c$, 
with respect to 
the massless case, which provide new ``conjugated'' coordinates; 
that is, extra degrees of freedom enter the theory through this 
pseudo-extension, which provides dynamics to the local (gauge) 
coordinates $\varphi_a$ of the structure subgroup $\TT$, dynamics which 
is transferred to the vector potentials $A_a$ to conform massive vector 
bosons. 

To understand fully the last statement concerning the interplay among 
different cocycles and mainly between the massless and massive cases, 
we must construct the Hilbert space of both theories explicitly. Let us 
proceed with the massless case and leave the peculiarities of the massive
one to the next section.  

The representation 
$L_{\tg'}\Psi(\tg)=\Psi(\tg'*\tg)$ of $\TG$ on $\TT$-equivariant wave 
functions (\ref{tequiv}) proves to be reducible. 
The reduction can be achieved by means of those right conditions 
(``polarization conditions'' \cite{GAQ}) 
$R_{\tg_p}\Psi(\tg')=\Psi(\tg'*\tg_p)\equiv\Psi(\tg')\,\,\forall \tg'\in\TG$ 
compatible with the $\TT$-equivariant  conditions (\ref{tequiv}), 
in particular with $\Xi\Psi=i\Psi$. In general, polarization conditions 
contain finite right-transformations generated by left-invariant 
vector fields $\xl{}$ devoid of dynamical content (that is, without a
conjugated counterpart), and half of the left-invariant vector 
fields related to dynamical coordinates (either ``positions'' or ``momenta''). 
The left-invariant vector fields without conjugated counterpart 
are the combinations 
\be
{\cal G}_c\equiv <\,\xl{\theta_a}
\equiv\xl{\varphi_a}-\f{1}{r}\nabla\cdot\xl{A_a},\,\, /
\,\, C^{ab}_c\lambda^c=0\,\,\forall b\,>\,.\label{afterpol}
\ee 
The {\it characteristic subalgebra} ${\cal G}_c$ can be completed to a 
{\it full polarization subalgebra} ${\cal G}_p$ in two different ways:
\be
{\cal G}_p^{(A)}\equiv <\,\xl{\theta_a}\in {\cal G}_c,\,\,\xl{A_b}\,\,
\forall b\,>,\;\;\;
{\cal G}_p^{(E)}\equiv <\,\xl{\theta_a}\in {\cal G}_c,\,\,\xl{E_b}\,\,
\forall b\,>,
\ee
each one giving rise to a different representation space: a) the electric 
field representation and b) the magnetic field representation, respectively.\\

\ni {\bf a) The electric field representation $\Psi_A$}.\\

The solution to the polarization conditions 
$R_{\tp}\Psi_A(\tg)=\Psi_A(\tg)\,,\,\,\forall \tp\in {G}_p^{(A)}\,,\,\,
\forall \tg\in \TG$ or, in infinitesimal form $\xl{}\Psi_A=0,\,\,\forall 
\xl{}\in {\cal G}_p^{(A)}$, proves to be:

\be
\Psi_A(A,E,U;\zeta)=\zeta e^{-\coci{AE-U\nabla U^{-1}E}}
\Phi_A(E)\,,\label{polaA}
\ee
\ni where $\Phi_A$ is an arbitrary functional of $E$. The left-action 
of a general element $\tg'=(A',E',U';\zeta')$ of $\TG$ on wave functions
$\Psi_A$ is:
\bea
L_{\tg'}\Psi_A(\tg)&=&\zeta'\zeta e^{-2\coci{A'U'E{U'}^{-1}+{U'}^{-1}
\nabla U'E+\um A'E'}}\nn\\
& & \cdot e^{-\coci{AE-U\nabla U^{-1}E}}
\Phi_A(E'+U'E{U'}^{-1})\,.\label{actionA}
\eea
\ni The particular case of $\tg'=\tk'=(0,0,U';1)\in T\subset \TT$ gives us the 
expression of the rest of $\TT$-equivariant conditions (\ref{tequiv}), 
i.e. the {\it constraint equations}:
\be
L_{\tk'}\Psi_A(\tg)=D_{\TT}^{(\epsilon)}(\tk')\Psi_A(\tg)\Rightarrow 
\Phi_A(E)=e^{-2\coci{{U'}^{-1}\nabla U'E}}
\Phi_A(U'E{U'}^{-1})\,,\label{ligaA}
\ee
\ni where we have chosen the trivial representation $D_{\TT}^{(\epsilon)}=1$ 
for $T$ (see below for more general cases). 

The polarized, $\TT$-equivariant wave functions (\ref{polaA},\ref{ligaA}) 
define the constrained Hilbert space ${\cal H}(\TG)$ of the theory, and the 
infinitesimal form $\xr{\tg'}\Psi$ of the finite left-action 
$L_{\tg'}\Psi(\tg)$ of $\TG$ on ${\cal H}(\TG)$ provides the action of the 
operators $\hat{A}_a,\hat{E}_a,\hat{G}_a$ on wave functions. 
Thus, the group $\TG$ is irreducibly 
and unitarily represented with respect to the natural scalar product 
$\langle\Psi|\Psi'\rangle=\int_{\TG}\mu(\tg)\Psi^*(\tg)\Psi'(\tg)$, where 
$\mu(\tg)$ denotes the standard left-invariant measure of $\TG$ [exterior 
product of the components of the left-invariant 1-form $\theta^L$].

The infinitesimal form of the finite expressions (\ref{actionA}) is:
\bea
\xr{A_a}\Psi_A&=&\zeta e^{-\coci{AE-U\nabla U^{-1}E}}iE_a
\Phi_A(E)\Rightarrow\hat{E}_a\Phi_A(E)=-E_a\Phi_A(E)\nn\\
\xr{E_a}\Psi_A&=&\zeta e^{-\coci{AE-U\nabla U^{-1}E}}\deriv{E_a}
\Phi_A(E)\Rightarrow \hat{A}_a\Phi_A(E)=i\deriv{E_a}\Phi_A(E)\nn\\ 
\xr{\varphi_a}\Psi_A&=&
\zeta e^{-\coci{AE-U\nabla U^{-1}E}}\left(-\f{i}{r}
\nabla\cdot E_a+C^{ab}_cE_b\cdot\deriv{E_c}\right)\Phi_A(E) \\
 &\Rightarrow& \hat{G}_a\Phi_A(E)=\left(-\f{1}{r}\nabla\cdot\hat{E}_a
-C^{ab}_c\hat{E}_b\cdot\hat{A}_c\right)\Phi_A(E)\,,\nn
\eea
\ni which provides the explicit expression for the basic operators of 
the theory. Several attempts \cite{Bauer} have been made to simplify the 
Gauss law constraint (\ref{ligaA}), which in infinitesimal form reads 
$\hat{G}_a(x)\Phi_A(E)=0$, by means of a unitary transformation 
$\Phi'_A(E)=\exp\left\{-\f{i}{r}\Omega(E)\right\}\Phi_A(E)$ in the electric 
field representation. The variation $\omega^a_j(E)=-\f{\p \Omega(E)}{\p
E_a^j}$ transforms as a standard Lie-algebra valued 
connection and modifies the operator 
$\hat{G}_a(x)$ so that the new constraint equations 
$\hat{G}'_a(x)\Phi'_A(E)=iC^{ab}_cE_b\cdot\deriv{E_c}\Phi'_A(E)=0$ reduce
to simple ``s-wave" conditions.\\

\ni {\bf b) The magnetic field representation $\Psi_E$}.\\ 

The choice of the polarization subalgebra ${\cal G}_p^{(E)}$ 
results in polarized wave functions of the form:
\be
\Psi_E(A,E,U;\zeta)=\zeta e^{\coci{AE-U\nabla U^{-1}E}}
\Phi_E(A+\nabla UU^{-1})\,,
\ee
\ni where $\Phi_E$ is an arbitrary functional of ${\cal A}\equiv
A+\nabla UU^{-1}$. The left-action of $\TG$ on wave functions $\Psi_E$ 
is now:
\bea
L_{\tg'}\Psi_E(\tg)&=&\zeta'\zeta e^{-2\coci{U'A{U'}^{-1}E'+{U'}^{-1}
E'U'\nabla UU^{-1}+\um A'E'-\um U'\nabla{U'}^{-1}E'}}\nn\\
& & \cdot e^{-\coci{AE-U\nabla U^{-1}E}}
\Phi_E\left(A'+\nabla U'{U'}^{-1}+U'(A+\nabla UU^{-1}){U'}^{-1}
\right)\,.\label{actionE}
\eea
\ni The constraint equations (\ref{ligaA})  in the present magnetic 
representation are: 

\be
L_{\tk'}\Psi(\tg)=D_{\TT}^{(\epsilon)}(\tk')\Psi(\tg)\Rightarrow 
\Phi_E({\cal A})=\Phi_E(U'{\cal A}{U'}^{-1}+\nabla U'{U'}^{-1})\,\label{ligaE}
\ee
\ni  [note the absence of a phase in comparison with the electric 
representation case]. 
The infinitesimal form of the finite expression (\ref{actionE}) is:
\bea
\xr{A_a}\Psi_E&=&\zeta e^{\coci{AE-U\nabla U^{-1}E}}\deriv{{\cal A}_a}
\Phi_E({\cal A})\Rightarrow \hat{E}_a\Phi_E({\cal A})= i\deriv{{\cal
A}_a}\Phi_E({\cal A})\nn\\
\xr{E_a}\Psi_E&=&-i\zeta e^{\coci{AE-U\nabla U^{-1}E}}{\cal A}_a
\Phi_E({\cal A})\Rightarrow \hat{A}_a\Phi_E({\cal A})=
{\cal A}_a\Phi_E({\cal A})\\
\xr{\varphi_a}\Psi_E&=&\zeta e^{\coci{AE-U\nabla U^{-1}E}}\left(-\f{1}{r}
\nabla\cdot \deriv{{\cal A}_a} +C^{ab}_c{\cal A}_b\cdot
\deriv{{\cal A}_c}\right)\Phi_E({\cal A})\,.\nn
\eea

Since $\TT$-equivariant conditions 
(\ref{tequiv},\ref{ligaA},\ref{ligaE}) are imposed as finite 
left-restrictions, it is 
evident that not all the operators $\xr{}$ will preserve the constraints; 
we shall call $\tilde{{\cal G}}_{{\tiny {\rm good}}}$ the subalgebra of 
({\it good}$\sim$physical) operators which will do so. These 
must be found inside the right-enveloping algebra 
${\cal U}(\tilde{{\cal G}}^R)$ 
of polynomials of the basic operators $\hat{A}_a(x),\,\hat{E}_b(x)$, as 
forming part of the {\it normalizer} of $T$; for example, 
 a sufficient condition for $\tilde{{\cal G}}_{\small{\rm good}}$ 
to preserve the constraints is $[\tilde{{\cal G}}_{\small{\rm good}},
\tilde{{\cal T}}]\subset {\rm Ker}\,\, d D_{\TT}^{(\epsilon)}$. 
In particular, some good operators are:
\be
\tilde{{\cal G}}_{{\rm {\small good}}}=<\,\tr\left[ 
\hat{E}^j(x)\hat{B}^k(x)\right],\,\tr\left[ 
\hat{E}^j(x)\hat{E}^k(x)\right],\, \tr\left[ \hat{B}^j(x)\hat{B}^k(x)\right],\,
\Xi\,>\,,\label{good}
\ee
\ni  where $\hat{B}_a\equiv \nabla\wedge\hat{A}_a-\um r C^{ab}_c
 \hat{A}_b\wedge\hat{A}_c$ (the magnetic field)
can be interpreted as a ``correction'' to $\hat{A}_a$   
that, unlike $\hat{A}_a$, transforms homogeneously under the adjoint 
action of $T$ [see 2nd line of (\ref{commutators})]. The components 
$\hat{\Theta}^{\mu\nu}(x)$ of the 
standard canonical energy-momentum tensor for Yang-Mills theories  
are linear combinations of operators in (\ref{good}); for example,  
$\hat{\Theta}^{00}(x)=-\tr\left[ \hat{E}^2(x)+\hat{B}^2(x)\right]$ 
is the Hamiltonian 
density. In this way, Poincar\'e invariance is retrieved in the 
constrained theory. At this stage, it is worth
mentioning that ${\cal G}_c$ would have included the entire Poincar\'e 
algebra had we incorporated the Poincar\'e group into $\TG$ 
(see \cite{empro,gtp} for the 
Abelian case). However, unlike other standard approaches 
to Quantum Mechanics, GAQ still remainss even in the absence of a 
well-defined (space-)time evolution, an interesting and desirable 
property concerning the quantization of gravity (see, for example, 
\cite{Rovelli2}).

Let us mention, for the sake of completeness, that the actual use
of good operators is not restricted to first- and second-order operators. 
Higher-order operators can constitute a useful tool in finding the whole 
constrained Hilbert space ${\cal H}_{{\small{\rm phys}}}(\TG)$. In fact, it 
can be obtained from a 
$\TT$-equivariant (physical) state $\Phi^{(0)}$, i.e. $\hat{G}_a\Phi^{(0)}=0$,
on which the energy-momentum tensor has null expectation value 
$\langle \Phi^{(0)}|\hat{\Theta}^{\mu\nu}|\Phi^{(0)}\rangle=0$, by taking the 
orbit of the rest of good operators passing through this ``vacuum''. This has 
indeed been a rather standard technique (the Verma module approach) in 
theories where null vector states are present in the original Hilbert space 
\cite{Kac,Feigin,virazorro}. From another point of view, with regard to  
confinement, exponentials of the form $\varepsilon_{\Sigma_2}
\equiv\tr\left[\exp(\epsilon_{jkl}\int_{\Sigma_2}{d\sigma^{jk}\hat{E}^l})
\right]$ and $\beta_{\Sigma_2}\equiv\tr\left[\exp(\epsilon_{jkl}
\int_{\Sigma_2}{d\sigma^{jk}\hat{B}^l})\right]$, 
where ${\Sigma_2}$ is a two-dimensional surface in three-dimensional space, 
are good operators related to Wilson loops.

As a step prior to tackling the massive case, 
let us show how new physics can enter the theory by considering 
non-trivial representations $D_{\TT}^{(\epsilon)}$ of $\TT$ or, in an 
equivalent way, by introducing certain extra pseudo-cocycles in the 
group law (\ref{ley}).

\subsection{$\vartheta$-Angle}
 
More general representations 
for the  constraint 
subgroup $T$, namely the one-dimensional representation 
$D_{\TT}^{(\epsilon)}(U)=e^{i\epsilon_{{}_U}}$,  
can be considered if we impose 
additional boundary conditions such as $U(x)
\stackrel{x\rightarrow\infty}{\longrightarrow}\pm I$; this means that we 
compactify the space $\Re^3\rightarrow S^3$, so that the group $T$ 
falls into disjoint homotopy classes $\{U_l\,,\,\epsilon_{{}_{U_l}}=
l\vartheta\}$ 
labeled by integers $l\in Z=\pi_3({\bf T})$ (the third homotopy group). 
The index $\vartheta$ (the {\it $\vartheta$-angle} 
\cite{Jackiwtheta}) parametrizes 
{\it non-equivalent quantizations}, in the same way 
that Bloch momentum $\epsilon$ does  
for particles in periodic potentials, where the wave function acquires 
a phase $\psi(q+2\pi)=e^{i\epsilon}\psi(q)$ after a translation of, 
let us say, $2\pi$. 
The phenomenon of non-equivalent quantizations can also be reproduced by 
keeping the  constraint condition  $D_{\TT}^{(\epsilon)}(U)=1$, as in 
(\ref{ligaA},\ref{ligaE}),  at the expense of introducing a new cocycle 
(indeed a coboundary)  $\xi_\vartheta$ which is added to the 
previous cocycle $\xi$ in (\ref{ley}). The generating function 
of $\xi_\vartheta$ is $\eta_\vartheta(g)=\vartheta\int{d^3x\, 
{\cal C}^0(x)}$, where ${\cal C}^0$ is 
the time component of the 
{\it Chern-Simons secondary characteristic class}
\be
{\cal C}^\mu=-\frac{1}{16\pi^2}\epsilon^{\mu\alpha\beta\gamma}{\rm tr}
({\cal F}_{\alpha\beta}{\cal A}_\gamma-\frac{2}{3}{\cal A}_\alpha 
{\cal A}_\beta {\cal A}_\gamma)\,,
\ee
which is the vector whose divergence equals the Pontryagin density  
 ${\cal P} = \partial_\mu {\cal C}^\mu = -\frac{1}{16\pi^2} 
{\rm tr} \break ({}^*{\cal F}^{\mu\nu} {\cal F}_{\mu\nu})$ 
(see \cite{Jackiw}, for instance). 
Like some total derivatives (namely, the Pontryagin density), 
which do not modify 
the classical equations of motion when added to the Lagrangian but have a 
non-trivial effect in the quantum theory, the coboundary   
$\xi_\vartheta$ gives rise to non-equivalent quantizations parametrized 
by $\vartheta$ when 
the topology of the space is affected by the imposition of 
certain boundary conditions (``compactification of the space''), 
even though it is a trivial cocycle of  the ``unconstrained'' theory. 
The phenomenon of non-equivalent quantizations can sometimes also be  
understood as an {\it Aharonov-Bohm-like effect} (an effect experienced by the 
quantum particle but not by the classical one) 
and the gradient $d\eta(g)$ can also be understood 
as an {\it induced gauge connection} (see e.g. \cite{Landsman,McMullan}, and 
\cite{FracHall} for the example 
of a superconducting ring threaded by a magnetic flux) which modifies  
momenta according to the minimal coupling. For our case, the induced 
gauge connection $\delta\eta_\vartheta(g)=\frac{\vartheta r^2}{8\pi^2}
B^a_j\delta A^j_a$ ($B^a_j$ is the magnetic field) modifies the momentum 
operators $\hat{E}_a\equiv i\xr{A_a}\rightarrow \hat{E}_a +
\frac{\vartheta r^2}{8\pi^2}\hat{B}_a$ and, accordingly, the Schr\"odinger 
equation $\int{d^3x\hat{\Theta}^{00}(x)}\Phi={\cal E}\Phi$ for 
stationary solutions $\Phi$ with energy ${\cal E}$. As is well 
known, the theory also exhibits a band energy structure of the 
form $\alpha+\beta\cos\vartheta$, the ground-state band 
functional $|\vartheta\rangle=\sum_l e^{il\vartheta}|0_l\rangle$ being a 
superposition of wave functionals $\Psi_l(A)=\langle A|0_l\rangle$ peaked 
near the classical zero-energy configurations (pure gauge potentials) 
$A_{(l)}=U_l\nabla U_l^{-1}$.

As already discussed, only coboundaries 
generated by functions $\eta(g)$ with non-trivial gradient 
$\left.\delta\eta(g)\right|_{g=e}\not=0$ at the identity $g=e$ 
(i.e. pseudo-cocycles),  namely $\xi_3=\xi_\lambda$,  
will provide a contribution to the connection form of the theory 
(\ref{thetagen}) and the 
structure constants of the original Lie algebra. However, as we have just 
seen, a coboundary generated by a global function on the original 
(infinite-dimensional) group $G$ having trivial gradient at the identity, 
namely $\xi_\vartheta$, can contribute the quantization 
with global (topological) effects as the new group has a 
non-equivalent global multiplication law. 

In both cases, non-trivial gauge transformation properties,  
$D_{\TT}^{(\epsilon)}(U)\not=1$, of the wave functional $\Phi(A)$ can be 
reproduced, as already mentioned,  by keeping the trivial 
representation $D_{\TT}^{(\epsilon)}(U)=1$ at the expense of introducing 
new (pseudo-) cocycles, $\xi_\vartheta$ or $\xi_\lambda$, in the 
centrally extended group law (\ref{ley}). 
However, whereas $\xi_\vartheta$ does not introduce new degrees of 
freedom into the theory, pseudo-cocycles such as $\xi_\lambda$ provide 
new couples of conjugated field operators, thus substantially
modifying  the theory. Let us examine this in more detail.

\section{The massive case: `spontaneous' symmetry `breaking' 
and  alternatives to the Higgs mechanism\label{ymc}}

The effect of the pseudo-cocycle $\xi_3\equiv\xi_\lambda$ for 
$\lambda\not=0$ is 
equivalent to inducing {\it internal} (`spinor-like') infinite-dimensional 
non-Abelian representations 
$D_{\TT}^{(\lambda)}$ of $\TT$. It modifies the commutation relations 
(\ref{commutators}) and the number of 
field degrees of freedom of the theory by  restricting the number 
of vector fields in the characteristic subalgebra ${\cal G}_c$ with 
respect to the massless case, where ${\cal G}_c\sim {\cal T}$. That is, new 
couples of generators $(\xr{\varphi_a},\xr{\varphi_b})$,  
with $C^{ab}_c\lambda^c\not=0$, become 
conjugated [see the last commutator of (\ref{commutators})] and, 
therefore, new basic operators enter the theory. To count the number 
of field degrees of freedom for a given structure subgroup $\TT$ and a given 
mass matrix $\lambda=\lambda_aT^a$, let us denote by 
$\tau={\rm dim}({\bf T})$ and $c={\rm dim}({\bf G}_c)$ the dimensions of the 
rigid subgroups of $T$ and $G_c$; in general, for an arbitrary mass matrix 
$\lambda$, we have $c\leq\tau$. Unpolarized, $U(1)$-equivariant functions 
$\Psi(A^j_a,E^j_a,\varphi_a)$ depend on $n=2\times 3\tau+\tau$ field 
coordinates in $d=3$ dimensions; polarization equations introduce $p=c+
\frac{n-c}{2}$ independent restrictions on 
wave functions, corresponding to $c$ non-dynamical coordinates in $G_c$ and 
half of the dynamical ones; finally, constraints impose 
$q=c+\frac{\tau-c}{2}$ additional restrictions which leave 
$f=n-p-q=3\tau-c$ field degrees of freedom (in $d=3$). 
Indeed, for the massive case, constraints are 
{\it second-class} and we can impose only a polarization subalgebra 
${\cal T}_p\subset\tilde{{\cal T}}$, which contains a 
characteristic subalgebra 
${\cal T}_c=<\,\xr{\varphi_a},\,\,{\rm with}\,\, C^{ab}_c
\lambda^c=0\,\,\forall b\,>
\subset \tilde{{\cal T}}$ (which is isomorphic to ${\cal G}_c$) 
and half of the rest of generators 
in $\tilde{{\cal T}}$ (excluding $\Xi$);\footnote{A similar situation 
happens in the bosonic string theory, where we can impose as constraints 
half of the Virasoro operators (the positive modes $\hat{L}_{n\geq 0}$) 
only; that is, the appearance of central terms in the Lie algebra 
(\ref{viral}) precludes the whole Virasoro algebra to be imposed as 
constraints, and only a polarization subalgebra can be consistently 
imposed.} In total,   
$q=c+\frac{\tau-c}{2}\leq \tau$ independent 
constraints, which lead to constrained wave functions having support on 
$f_{m\not=0}=3\tau-c\geq f_{m=0}$ arbitrary fields; these fiels correspond to 
$c$ massless vector bosons attached to ${\cal T}_c$ and 
$\tau-c$ massive vector bosons.  In particular, for the massless 
case, we have ${\cal T}_c={\cal T}$, i.e. $c=\tau$,  
since constraints are {\it first-class} 
(that is, we can impose $q=\tau$ restrictions) and constrained wave 
functions have support on $f_{m=0}=3\tau-\tau=2\tau\leq f_{m\not=0}$ 
arbitrary fields corresponding to $\tau$ massless vector bosons. 
The subalgebra ${\cal T}_c$ corresponds to the unbroken 
gauge symmetry of the constrained theory and proves to be an  
{\it ideal} of $\tilde{{\cal G}}_{\small{\rm good}}$ [remember 
the characterization of {\it good} operators before Eq. (\ref{good}); see 
also Refs. \cite{config,gtp} for a definition and subtle distinctions 
between constraints and gauge symmetries inside GAQ]. 

Let us work out a couple of examples. Cartan (maximal Abelian) 
subalgebras of ${\bf T}$ will be preferred as candidates for the rigid 
subgroup of the unbroken electromagnetic gauge symmetry. Thus, let us use the 
Cartan basis $<\,H_i,E_{\pm \alpha}\,>$ instead of $<\,T^a\,>$, and denote 
$\{\varphi_i,\varphi_{\pm \alpha}\}$ the coordinates of $T$ 
attached to this basis (i.e. $\varphi_{\pm \alpha}$ are complex field 
coordinates attached to each root $\pm\alpha$, and $\varphi_i$ are 
real field coordinates attached to the maximal torus of {\bf T}). 
For $T=SU(2)(x)$ and $\lambda=\lambda_1H_1$, the 
characteristic, polarization and constraint subalgebras (leading 
to the electric field representation) are:
\be
{\cal G}_c=<\,\xl{\theta_1}\,>,\,\,\,
{\cal G}_p^{(A)}=<\,\xl{\theta_1},\xl{\theta_{+1}},\xl{A}\,>,\,\,\,
{\cal T}_p=<\,\xr{\varphi_1},\xr{\varphi_{-1}}\,>\,.
\ee
Indeed, the appearance of a central term in the commutator 
\be
\l[ \xr{\varphi_{+1}(x)}, \xr{\varphi_{-1}(y)}\r]= 
i\delta(x-y)\xr{\varphi_1(x)}+i\frac{\lambda_1}{r^2}\delta(x-y)\Xi
\ee
prevents the vector fields  $\xr{\theta_{\pm 1}}$  from  
being in ${\cal G}_c$ and precludes the simultaneous imposition of 
$\xr{\varphi_{-1}}\Psi_{{\small{\rm phys.}}}=0$ and 
$\xr{\varphi_{+1}}\Psi_{{\small{\rm phys.}}}=0$ as constraints 
(for the trivial representation $D_{\TT}^{(\epsilon)}(U)=1$), so that 
a polarization subalgebra ${\cal T}_p$ is the only option (${\cal T}_p$ 
has to contain the `negative modes' $\xr{\varphi_{-1}}$ when  
the `positive' ones 
$\xl{\theta_{+1}}$ have been chosen in ${\cal G}_p^{(A)}$, 
or the other way round). The new couple of 
{\it basic} operators $\hat{G}_{\pm 1}\equiv \xr{\varphi_{\pm 1}}$ 
(these are basic because they can no longer be written 
in terms of $\hat{A}$ and $\hat{E}$) 
represent two new field degrees of freedom which are transferred to the 
vector potentials  $\hat{A}_{\pm 1}$ to conform massive vector bosons; i.e. 
$\hat{G}_{\pm 1}$ can be seen as the longitudinal component of 
$\hat{A}_{\pm 1}$, which is missing (is zero) in the massless case. Thus,
the constrained theory corresponds to a self-interacting field 
theory of a massless vector 
boson $A_1$ with `unbroken' gauge subgroup $T_c=U(1)(x)\subset SU(2)(x)$ and 
two charged vector bosons $A_{\pm 1}$ with mass cubed $m^3_1=\lambda_1$.  

For $T=SU(3)(x)$ and $\lambda=\lambda_2 H_2$, we have
\be\ba{cc}
{\cal G}_c=<\,\xl{\theta_{1,2}},\xl{\theta_{\pm 1}}\,>,\,\,\,
{\cal G}_p^{(A)}=<\,\xl{\theta_{1,2}},\xl{\theta_{\pm 1}},
\xl{\theta_{+2,+3}},\xl{A}\,>,\\ 
{\cal T}_p=<\,\xr{\varphi_{1,2}},\xr{\varphi_{\pm 1}},
\xr{\varphi_{-2,-3}}\,>\ea\,.
\ee
Indeed, in this case, the relevant commutators 
\bea
\l[ \xr{\varphi_{+1}(x)}, \xr{\varphi_{-1}(y)}\r]&=& 
i\delta(x-y)\xr{\varphi_1(x)}\,,\nn\\
\l[ \xr{\varphi_{+2}(x)}, \xr{\varphi_{-2}(y)}\r]&=& 
\frac{i}{\sqrt{3}}\delta(x-y)\xr{\varphi_1(x)}+i\delta(x-y)\xr{\varphi_2(x)}+
i\frac{\lambda_2}{r^2}\delta(x-y)\Xi\,,\label{su3}\\
\l[ \xr{\varphi_{+3}(x)}, \xr{\varphi_{-3}(y)}\r]&=& 
\frac{-i}{\sqrt{3}}\delta(x-y)\xr{\varphi_1(x)}+i\delta(x-y)\xr{\varphi_2(x)}+
i\frac{\lambda_2}{r^2}\delta(x-y)\Xi\,,\nn
\eea
reveal that the vector fields  $\xr{\theta_{\pm 2}}$ and 
$\xr{\theta_{\pm 3}}$ have dynamical content and cannot be included in 
${\cal G}_c$. Also, its conjugated character precludes the simultaneous 
imposition of $\xr{\varphi_{-2,-3}}$ and $\xr{\varphi_{+2,+3}}$ 
as constraints, and a polarization subalgebra ${\cal T}_p$ has to be 
chosen. On the contrary, the vector fields  $\xr{\varphi_{\pm 1}}$ are 
devoid of dynamical content, as can be seen from the first line of 
(\ref{su3}), and can be simultaneously imposed as constraints in 
${\cal T}_p$ (this is because of the particular choice of mass matrix 
$\lambda$, which determines different ``symmetry breaking'' patterns).  
As for $T=SU(2)(x)$, the new couples of 
basic operators $\hat{G}_{\pm 2,\pm 3}\equiv \xr{\varphi_{\pm 2,\pm 3}}$ 
represent four new field degrees of freedom which are transferred to the 
vector potentials  $\hat{A}_{\pm 2,\pm 3}$ to conform massive vector bosons. 
Thus, the constrained theory corresponds to a 
self-interacting theory of two massless vector 
bosons $A_{1,2}$, two massless charged vector bosons $A_{\pm 1}$ 
[the `unbroken' gauge subgroup is now $T_c=SU(2)\times U(1)(x)\subset 
SU(3)(x)$] and four charged vector bosons $A_{\pm 2,\pm 3}$ with mass cubed 
$m^3_2=\lambda_2$. 

Summarizing, new basic operators 
$\hat{G}_{\pm \alpha}\equiv \xr{\varphi_{\pm\alpha}}$,  
with $C^{\alpha-\alpha}_i\lambda^i\not=0$,  and new   
good operators $\hat{C}_i=\{$Casimir operators of $\TT\}$ 
($i$ runs the range of {\bf T}) enter the theory, in
contrast to the massless case. For example, for $T=SU(2)(x)$, the Casimir 
operator is 
\be
\hat{C}(x)= (\hat{G}_1(x)+\frac{\lambda_1}{r^2})^2
+2(\hat{G}_{+1}(x)\hat{G}_{-1}(x)
+\hat{G}_{-1}(x)\hat{G}_{+1}(x))\,.
\ee 
Also, the Hamiltonian density 
$\hat{\Theta}^{00}(x)=-\tr\left[ E^2(x)+B^2(x)\right]$ for $m=0$ can be   
affected in the massive case $m\not=0$ by the presence of extra terms 
proportional to these Casimir operators  as follows:
\be
\hat{\Theta}^{00}_{m\not=0}(x)=\hat{\Theta}_{m=0}^{00}(x)+\sum_i{
\frac{r^2}{m_i^2}\hat{C}_i(x)}\,.
\ee
Thus, the Sch\"odinger equation 
$\int{d^3x\hat{\Theta}^{00}_{m\not=0}(x)}\Phi={\cal E}\Phi$ is also 
modified by the presence of extra terms.

As already mentioned in reference to the Virasoro group, 
pseudo-cocycle parameters such as $\lambda_i$ are usually hidden 
in a redefinition of the generators 
involved in the pseudo-extension 
$\hat{G}_i(x)+\lambda_i/r^2\equiv \hat{G}_i'(x)$. However, whereas the 
vacuum expectation value $\langle 0_\lambda|\hat{G}_i(x)|0_\lambda\rangle$ is 
zero,\footnote{it can be easily proven taking into account that the vacuum is 
annihilated by the right version of the polarization subalgebra 
dual to ${\cal G}_p$ \cite{conforme}; also, $\hat{G}_i=\xr{\varphi_i}$ is 
always in ${\cal T}_p$; that is, it is zero on constrained wave functionals 
$\Psi_{{\small{\rm phys.}}}$, including the physical vacuum.} the vacuum 
expectation value $\langle 0_\lambda|\hat{G}_i'(x)|0_\lambda\rangle=
\lambda_i/r^2$ 
of the redefined operators $\hat{G}_i'(x)$ is non-null and proportional 
to the cubed mass in the `direction' $i$ of the `unbroken' gauge 
symmetry $T_c$, which depends on the particular choice of 
the mass matrix $\lambda$. Thus, the effect of the pseudo-extension 
manifests also in a different choice of a vacuum in which some 
gauge operators have a non-zero expectation value. 
This fact reminds us of the Higgs mechanism 
in non-Abelian gauge theories, where the Higgs fields point to the direction 
of the non-null vacuum expectation values. However, the spirit of the  
Higgs mechanism, as an approach to supply mass, and the one 
discussed in this paper are 
radically different, even though they have some common characteristics. 
In fact, we are not making use of extra scalar fields in the theory to 
provide mass to the vector bosons, but it is the gauge group itself 
that acquires dynamics for the massive case and transfers degrees of freedom 
to the vector potentials.
 
Before finishing, let us show how to incorporate fermionic matter into 
the theory and outline the main changes in the foregoing discussion 
had we considered it from the beginning. 

\section{Incorporating fermionic matter\label{fermionic}}
 
Fermionic matter can enter the theory through extra 
(Dirac) field coordinates $\psi_l(x),\, l=1,\dots, p$, which we can 
assemble into a column vector $\psi(x)$, 
and an extra cocycle $\xi_{{\small{\rm matter}}}$  
leading to a quantizing {\it supergroup} $\widetilde{SG}$. The group law that 
describes this boson-fermion gauge theory is 
(\ref{ley}) together with 
\bea
\psi''(x)&=&\psi'(x)+\rho({U(x)})\psi(x)\,,\nn\\
\bar{\psi}''(x)&=&\bar{\psi}'(x)+\bar{\psi}(x)\rho({U(x)^{-1}})\,,\\
\xi_{{\small{\rm matter}}}&\equiv& i\int{{d}^3x
\,\left(\bar{\psi}' \gamma^0\rho(U')\psi-\bar{\psi}\rho(U'^{-1}) 
\gamma^0\psi'\right)}\,,\nn
\eea
where $\rho(U)$ is a $p$-dimensional representation of {\bf T} acting on 
the column vectors $\psi$, and $\gamma^0$ is the time component of 
the standard Dirac matrices $\gamma^\mu$. To compute the 
left- and right-invariant super-vector fields $\tilde{X}^{L,R}$ 
and the polarized super-wavefunctionals $\Psi(A,E,U,\psi,\bar{\psi};\zeta)$,  
we have to take into account the Grassmann character of the Dirac field 
coordinates. The unitary irreducible representations of $\widetilde{SG}$ 
can easily be constructed by following 
the main steps described 
in this article and by taking care of the subtleties introduced by 
the anti-commutation of Grassmann variables (see \cite{geo} for the 
finite-dimensional example of the super-Galilei group 
$\widetilde{SG}_{(m)}$). We should mention that, in the presence of 
fermion sources, the infinitesimal 
version of the constraint (\ref{ligaA}), i.e. the Gauss law, is modified to 
\be
\hat{G}_a\Phi_{A,\psi}(E,\bar{\psi})=\left(-\f{1}{r}\nabla\cdot\hat{E}_a
-C^{ab}_c\hat{E}_b\cdot\hat{A}_c-\frac{i}{r}\hat{\bar{\psi}}\gamma^0 
\tau_a\hat{\psi}\right)\Phi_{A,\psi}(E,\bar{\psi})=0\,,
\ee
(where $\tau_a$ denote the generators of $\rho$) in accordance with 
other standard approaches. Other interesting questions like chiral 
anomalies are left to future publications.

\section{Some comments and outlook\label{conclu}}

One question which is worthwhile to comment upon is the preservation of 
renormalizability for a non-trivial mass matrix $\lambda\not=0$. 
Since our approach to quantization is not perturbative, we must 
answer this question using general arguments. In fact, from 
a group-theoretical point of view, there is no reason why a given 
unitary irreducible representation of a group $\TG$ (namely, 
the massive one) can show bad properties, like 
`inescapable divergences', whereas other 
(namely, the massless one) does not. Even more, when we use the term 
`unbroken gauge symmetry', in referring to $T_c$, we mean simply the
subgroup of $\TT$ devoid of dynamical content; the gauge group 
of the constrained theory is, in both the massless and massive cases, 
the group $T=\TT/U(1)$, although, for the massive case, only a polarization 
subgroup $T_p$ can be consistently imposed as a constraint. This is also 
the case of the Virasoro algebra (\ref{viral}) in String Theory, where 
the appearance of central terms does not spoil gauge invariance but 
forces us to impose half of the 
Virasoro operators only (the positive modes $\hat{L}_{n\geq 0}$) 
as constraints.

Thus, the `spontaneous breakdown' of the gauge symmetry group $T$ manifests 
through non-trivial transformations of the phase $\zeta$ of the wave 
functional $\Psi$ under the action of $T$, leading to the appearance 
of new `internal' field degrees  of freedom which modify the 
`field mass content' of some vector potentials $A$, depending on the 
choice of mass-matrix elements $\lambda^i=-2\tr(T^i\lambda)$. This 
situation recalls the important physical implications of 
{\it geometric phases} (namely, Berry's phase) in quantum mechanics, 
the case discussed in the present paper being a particular one. 
In other words, 
the presence of mass is detected by the wave functional $\Psi$ in its 
`gauge excursions' through the configuration space, 
as happens with the presence of monopoles (see Ref. 
\cite{McMullan,Landsman} for 
a discussion on the emergence of gauge structures 
---the ``$H$-connection''--- and generalized spin 
when quantizing on a coset space $G/H$).  Also, the zeroes (critical 
values) of the mass-matrix elements $\lambda^i$ correspond to different 
phases of the physical system characterized by the corresponding 
unbroken gauge symmetry $T_c$; thus, the system can undergo `spontaneous' 
phase-transitions between different phases corresponding to 
non-equivalent fibrations $\TT$ of $T$ by $U(1)$ (i.e. different choices 
of characteristic subgroups $T_c$ of $\TT$).

Open questions remain 
about what happens when a ``true'' cocycle $\xi_3$ exists; 
for example,  we can find non-trivial central extensions $\TT$ 
of $T={\rm Map}(S^1,{\bf T})$ by $U(1)$ (Kac-Moody groups) in one compact 
spatial dimension, deformations which correspond 
to anomalous situations in the standard (canonical) 
approach to quantization of gauge theories. This fact makes the quantization 
of `massive' Yang-Mills fields (in this scheme) not so 
trivial, even in one 
spatial dimension. Also, it would be worth exploring the richness of the case 
${\bf T}=SU(\infty)$ (infinite number of colours), the Lie-algebra of which 
is related to the (infinite-dimensional) Lie-algebra of area preserving 
diffeomorphisms of the sphere $SDiff(S^2)$ (see \cite{Floratos} and 
references therein). In general, the cohomological richness, i.e. the number 
of inequivalent central (pseudo) extensions, of 
$T={\rm Map}(M,{\bf T})$ depends on the topology of $M$. Also, as  
usually happens with central charges, a quantization 
of the mass parameters $m_c\sim (n)^{1/3},\,\,n=1,2,3,\dots$ could arise 
from the compact character of the involved manifolds.

Another question that deserves further study is, of course, 
the physical implications that this new point of view carries along.

\appendix

\section{Appendix: a $0+1$D analogy}

This appendix is intended to clarify ideas by providing a simple, 
but illustrative, quantum mechanical analogy  
which contains most of the essential elements exposed in the paper. 
Indeed, a $SU(2)$ gauge invariant Yang-Mills theory in 
$0+1$D may  eventually be 
related to a spinning particle with constraints 
(zero total angular momentum) inside the present GAQ framework. 

Let us denote by $A\equiv r\left(\begin{array}{cc} A_0 & A_+ \\ A_- & -A_0
 \end{array}\right),\,A_0\equiv A_3,\,A_\pm \equiv A_1\pm iA_2,\,$ the 
$su(2)$-valued vector potentials, and let us choose the following set of 
coordinates 
\bea
e^{i\varphi_0} \equiv\f{z_1}{|z_1|}, \;\; \c\equiv\f{z_2}{z_1},\;\;
\cc\equiv\f{z_2^*}{z_1^*},\;\;\;\; e^{i\varphi_0}\in S^1;
\;\;\c,\cc\in S^2\,,\label{z}
\eea
for the gauge group  
\be
SU(2)\equiv\left\{ U= \left( \begin{array}{cc} z_1&z_2\\-z_2^*&z_1^*\end{array}
\right) ,z_i,z_i^* \in C/ \det(U)=|z_1|^2+|z_2|^2=1\right\}\,. 
\ee 
Let us also work in an holomorphic picture and define $\bos\equiv 
\f{1}{\sqrt{2}r}(A+iE)$ and $\bosc\equiv \f{1}{\sqrt{2}r}(A-iE)$. The adjoint 
action of the gauge group on the vector potential $A$ and the electric field 
$E$ can be explicitly written as
\be
U\bos U^{-1}=\f{1}{1+\c\cc}\left( \begin{array}{cc} 
e^{i\varphi_0} &\c e^{i\varphi_0} \\ -\cc e^{-i\varphi_0}& e^{-i\varphi_0}
\end{array}\right)
\left(\begin{array}{cc} \bos_0 & \bos_+ \\ \bos_-& -\bos_0
 \end{array}\right)\left( \begin{array}{cc} 
 e^{-i\varphi_0} &-\c e^{i\varphi_0} \\ \cc e^{-i\varphi_0}& e^{i\varphi_0}
\end{array}\right)\,,
\ee
and the centrally extended group law (\ref{ley}) now adopts the form
\bea
U''&=&U'U\,,\nn\\
 \bos''&=&  \bos +  U^{-1}\bos'U\,,\nn\\
\bosc''&=&  \bosc +  U^{-1}\bosc'U\,,\\
\zeta''&=&\zeta'\zeta\exp\f{1}{4}\tr\left[\left(\begin{array}{cc} \bos & 
\bosc \end{array}\right)S\left(\begin{array}{c} U^{-1}\bos'U 
\\ U^{-1}\bosc'U \end{array}\right)\right]\exp{2i\lambda
(\varphi_0''-\varphi_0'-\varphi_0)}\,,\nn
\eea
where we miss the mixed cocycle $\xi_2$ because we are working in $0$ 
spatial dimensions (we are restricting ourselves to a ``single point''). 
We are also keeping only the (relevant) linear term $\lambda\varphi_0$ 
in the expansion of $\tr[\lambda\sigma_3\log U]$ ($\sigma_3$ 
is the standard Pauli matrix). The left- and right-invariant vector 
fields are explicitly:
\bea
\xl{\zeta}&=&\xr{\zeta}=\zeta\f{\p}{\p\zeta}\,,\\
\xl{\bos} &=&\f{\p}{\p\bos}+\um\bosc\zeta\f{\p}{\p\zeta}\,,\;\;
\xl{\bosc} =\f{\p}{\p\bosc}-\um\bos\zeta\f{\p}{\p\zeta}\,,\nn\\
\xl{\varphi_0} &=&\f{\p}{\p\varphi_0}-2i\c\f{\p}{\p \c}+2i\cc
\f{\p}{\p \cc}-2(\bos\times\f{\p}{\p\bos})_0-
2(\bosc\times\f{\p}{\p\bosc})_0\,,\nn   \\
\xl{\c} &=&\f{-i}{2}\cc \f{\p}{\p\varphi_0}+\f{\p}{\p \c}+
\cc^2\f{\p}{\p \cc}
+i(\bos\times\f{\p}{\p\bos})_- + i(\bosc\times\f{\p}{\p\bosc})_-
+\lambda\cc\zeta\f{\p}{\p\zeta}\,,\nn \\
\xl{\cc} &=&\f{i}{2} \c \f{\p}{\p\varphi_0}+\c^2 \f{\p}{\p \c}+
\f{\p}{\p \cc}-i(\bos\times\f{\p}{\p\bos})_+ - i(\bosc\times\f{\p}{\p\bosc})_+ 
-\lambda\c\zeta\f{\p}{\p\zeta}\,,\nn\\ 
& & \nn\\
\xr{\bos}&=&U\f{\p}{\p\bos}U^{-1}
-\um U\bosc U^{-1}\zeta\f{\p}{\p\zeta}\,,\nn\\
\xr{\bosc}&=&U\f{\p}{\p\bosc}U^{-1}
+\um U\bos U^{-1}\zeta\f{\p}{\p\zeta}\,,\nn\\
\xr{\varphi_0} &=&\f{\p}{\p\varphi_0}\,, \nn \\
\xr{\c} &=& \f{i}{2} e^{-2i\varphi_0} \cc \f{\p}{\p\varphi_0}+
e^{-2i\varphi_0}(1+\c\cc)
\f{\p}{\p \c}-\lambda e^{-2i\varphi_0}\cc\zeta\f{\p}{\p\zeta}\,,\nn \\
\xr{\cc} &=& -\f{i}{2}e^{2i\varphi_0} \c \f{\p}{\p\varphi_0}
+e^{2i\varphi_0}(1+\c\cc)\f{\p}{\p\cc}
+\lambda e^{2i\varphi_0}\c\zeta\f{\p}{\p\zeta}\,,\nn
\eea
\ni where $(A\times B)_a\equiv \epsilon^{abc}A_bB_c,\,\epsilon^{123}=1$,  
denotes the vector product and $(A\times B)_\pm\equiv 
(A\times B)_1\pm i(A\times B)_2$. 
The  commutators (\ref{commutators}) now adopt the following form:
\be \ba{lll}
 \l[\xr{\bos_+},\xr{\bosc_-}\r]=
-i\Xi & \l[\xr{\bos_-},\xr{\bosc_+}\r]=
-i\Xi & \l[\xr{\bos_0},\xr{\bosc_0}\r]=-i\Xi   \\
\l[\xr{\varphi_0},\xr{\c}\r]=-2i\xr{\c}&\l[\xr{\varphi_0},\xr{\cc}\r]=
2i\xr{\cc} & \l[\xr{\c},\xr{\cc}\r]=-i\xr{\varphi_0}-2i\lambda\Xi \\
\l[\xr{\varphi_0},\xr{\bos_0}\r]=0 &
\l[\xr{\varphi_0},\xr{\bos_+}\r]=-2i\xr{\bos_+} &
\l[\xr{\varphi_0},\xr{\bos_-}\r]=2i\xr{\bos_-} \\
\l[\xr{\c},\xr{\bos_0}\r]=2\xr{\bos_+} & \l[\xr{\c},\xr{\bos_+}\r]=
0 & \l[\xr{\c},\xr{\bos_-}\r]=-\xr{\bos_0} \\
\l[\xr{\cc},\xr{\bos_0}\r]=2\xr{\bos_-} &\l[\xr{\cc},\xr{\bos_+}\r]=
-\xr{\bos_0} & \l[\xr{\cc},\xr{\bos_-}\r]=0 
\ea\label{algebra}
\ee
\ni where we have omitted the commutators 
$\l[\xr{\varphi_0,\c,\cc},\xr{\bosc_j}\r]$, which have the same form 
as for the $\xr{\bos_j}$ vector fields.  One can also work out easily the  
{\it Quantization 1-form} (\ref{thetagen}), which  is:
\be
\Theta = \f{i}{4}\tr[\dosc d\dos -\dos d\dosc ]+
\overbrace{\frac{i\lambda}{1+\ch\chc}(\chc d\ch-\ch d\chc )}^{\Theta_{SU(2)}}
-i\zeta^{-1}d\zeta\,,\label{tetatot}
\ee
\ni where we denote 
$\dos\equiv U\bos U^{-1},\,\dosc\equiv U\bosc U^{-1},\,\ch\equiv 
e^{2i\varphi_0}\c,\,\chc\equiv e^{-2i\varphi_0}\cc$. 
 The {\it characteristic subalgebra} is just
\be
{\cal G}_c=<\xl{\varphi_0} >\,,\label{char}
\ee 
\ni and a full-polarization subalgebra exists for arbitrary (non-zero) 
$\lambda$, which is:
\be
{\cal G}_p=<\xl{\varphi_0},\,\xl{\c},\,\xl{\bos}>\,.
\label{pola}
\ee
\ni The general solution to the polarization equations 
$\xl{}\Psi=0,\,\xl{}\in {\cal G}_p$ leads to 
a Hilbert space ${\cal H}^{(\lambda)}(\TG)$ of wave functions of the form:
\be
\Psi^{(\lambda)}(\zeta,\varphi_0,\c,\cc,\bos,\bosc)=\zeta(1+
\c\cc)^{-\lambda}
e^{-\f{1}{4}\tr [\bosc\bos]}\Phi(\chc,\dosc),\label{wavepol}
\ee
where $\Phi$ is an arbitrary power series in the variables $\chc$ and $\dosc$. 
A scalar product can be given through the invariant integration volume 
of $\TG$:
\be
\mu(\tg)=\f{-i}{(1+\c\cc)^2}
\l[\prod_{a=1}^3dA_a\wedge dE_a\r]
\wedge \l[ d\hbox{Re}(\c)\wedge d\hbox{Im}(\c)\wedge  
d\varphi_0\r]\wedge
 \zeta^{-1}d\zeta\,.\label{volucho}
\ee
The phase space related to this quantum system is clearly 
$\Re^3\times\Re^3\times S^2$, as can be inferred from the symplectic 
form $\omega\equiv d\Theta/{\cal G}_c$ (the quotient of $d\Theta$ 
by the trajectories generated by left-invariant vector fields 
in (\ref{char})), the parameter $\lambda$ being the analogous of the 
spin $s$.

The constraint equations 
\bea
\xr{\varphi_0}\Psi^{(\lambda)}_{{\small{\rm phys.}}}=0
&\Rightarrow& \chc\f{\p\Phi}{\p\chc}
+i(\dosc\times\f{\p\Phi}{\p\dosc})_0=0\,,\nn\\
\xr{\cc}\Psi^{(\lambda)}_{{\small{\rm phys.}}}=0
&\Rightarrow& \f{\p\Phi}{\p\chc}
+i(\dosc\times\f{\p\Phi}{\p\dosc})_+=0\,,\label{ceroang}
\eea
keep 2 degrees of freedom out of the original $4=3+1$ 
corresponding to this ``spinning-like 
particle''. They can be interpreted as zero total angular-momentum 
(orbital+spin) conditions. Note that the condition 
\be 
\xr{\c}\Psi^{(\lambda)}_{{\small{\rm phys.}}}=0
\Rightarrow -2\lambda\chc\Phi+\chc^2\f{\p\Phi}{\p\chc}
-i(\dosc\times\f{\p\Phi}{\p\dosc})_-=0\label{ceroang2}
\ee
is incompatible with both conditions in (\ref{ceroang}), 
which correspond to a polarization subalgebra 
${\cal T}_p=<\xr{\varphi_0},\,\xr{\cc}>$ of ${\cal T}$, unless $\lambda=0$.
For $\lambda=0$, the characteristic subalgebra 
(\ref{char}) contains the whole $su(2)$ subalgebra, $\Phi$ does no longer 
depend on $\chc$, and the constraint conditions (\ref{ceroang},\ref{ceroang2}) 
keep a ``radial'' dependence of $\Phi$ on $R^2\equiv\um\tr[\dos\dosc]$ 
(``s-waves''), as corresponds to a spin-zero particle with 
zero orbital angular momentum. 
 
The good operators are 
\be
\tilde{{\cal G}}_{{\rm {\small good}}}=<\,\tr[\hat{\bos}^2],\,
\tr[\hat{\bosc}^2],\,\tr[\hat{\bos}\hat{\bosc}],\,\hat{C},\,\Xi\,>\,,
\ee
where $\hat{C}=(\xr{\varphi_0}+2\lambda\Xi)^2+2\xr{\c}\xr{\cc}+2\xr{\cc}
\xr{\c}$ is the Casimir operator of $SU(2)$.

\section*{Acknowledgment}

M. Calixto would like to thank the University of Granada for a Post-doctoral 
grant and the Department of Physics of Swansea for its hospitality. \\

\ni {\bf Note added.} We thank the referee who brought 
the reference \cite{McMullan} to our 
attention. It contains a nice summary of 
a generalization of Dirac's method of quantization of constrained systems by 
using Mackey's theory of inequivalent quantizations on a coset space $G/H$. 
The reader may find it interesting to compare GAQ and this generalized 
version of Dirac's approach by using the simple example given 
in the Appendix. Both approaches share the idea of 
``emergence of new (internal) degrees of freedom, existence of inequivalent 
quantizations and the appearance of an $H$-connection'' when constraints 
become second class. In fact, the role played by the characteristic 
subgroup $G_c$ in GAQ is similar to the role played by $H$ when quantizing 
on a coset space $G/H$; also, the piece 
$\Theta_{SU(2)}=\left.\f{\p}{\p g^j}\xi(g'|g)_\lambda\right|_{g'=g^{-1}}dg^j$ 
of the general connection form (\ref{thetagen}) in Eq. (\ref{tetatot}) 
corresponds to a ``$SU(2)$-connection''. However, an important distinction 
has to be made between both schemes of constrained quantization. 
The counterpart of the constraint equations (right conditions) 
\[ R_h\Psi(g)=\Psi(g*h)\equiv\Psi(g)\,,\,\,\forall h\in H,\,g\in G\,, \]
in the generalized Dirac's approach to the constrained quantization on 
$G/H$, are the polarization equations of GAQ (see paragraph before Eq. 
(\ref{afterpol})) which, in contrast, are intended to {\it reduce} 
the (left) regular representation  
$L_{g'}\Psi(g)=\Psi(g'*g)$ of $G$ on wave functions $\Psi$. In brief, 
GAQ further ``constrains'' wave functions by means of {\it extra} 
$T$-equivariance conditions (\ref{tequiv}) like (\ref{ceroang}), which are 
not present in the generalized 
Dirac's scheme of quantization. Also, $T$-equivariance conditions in GAQ 
force the definition of {\it good operators} (observables), concept 
which is absent in the other scheme.

\end{document}